\documentclass[aps,floatfix,showpacs,twocolumn]{revtex4}
\usepackage{graphicx}
\def\emptyline{\vspace{12pt}}

\begin{document}

\title{An inverse approach to Einstein's equations for non-conducting fluids}
\author{Mustapha Ishak$^1$\footnote{Electronic address: mishak@princeton.edu} and Kayll Lake$^2$\footnote{Electronic address: lake@astro.queensu.ca}}
\affiliation{
$^1$Department of Physics, Princeton University, Princeton, NJ 08544, USA\\
$^2$Department of Physics, Queen's University, Kingston, Ontario, Canada, K7L 3N6 }
\date{\today}

\begin{abstract}
We show that a flow (timelike congruence) in any type $B_{1}$
warped product spacetime is uniquely and algorithmically
determined by the condition of zero flux. (Though restricted,
these spaces include many cases of interest.) The flow is written
out explicitly for canonical representations of the spacetimes.
With the flow determined, we explore an inverse approach to
Einstein's equations where a phenomenological fluid interpretation
of a spacetime follows directly from the metric irrespective of
the choice of coordinates. This approach is pursued for fluids
with anisotropic pressure and shear viscosity. In certain
degenerate cases this interpretation is shown to be generically
not unique. The framework developed allows the study of exact
solutions in any frame without transformations. We provide a
number of examples, in various coordinates, including spacetimes
with and without unique interpretations. The results and
algorithmic procedure developed are implemented as a computer
algebra program called GRSource.
\end{abstract}


\maketitle

\section{Introduction}
The usual procedure for finding exact solutions of Einstein's
equations involves writing down a phenomenological energy-momentum
tensor, often a perfect fluid, in a set of coordinates, frequently
comoving, so that the field equations can be integrated, often
with the aide of simplifying assumptions \cite{kshm}
\cite{krasinski}. In view of the difficulty of solving Einstein's
equations, inverse problems are important \cite{solutions}. An
inverse problem of interest (not involving  the classification
problem for the Ricci tensor \cite{class} explicitly) may be
stated thus: Given a spacetime $(\mathcal{M},\bf{g})$ with
manifold $\mathcal{M}$  and Lorentzian metric $\bf{g}$, what, if
any, fluid flow could generate $(\mathcal{M},\bf{g})$ via
Einstein's equations? This question is explored here for the case
of type $B_{1}$ warped product spacetimes (see below). The problem
is of interest since, as is shown, the flow (velocity field) is
uniquely determined subject to a zero flux condition. With the
flow determined, physical parameters, subject to a fluid
decomposition that includes anisotropic pressure and shear
viscosity, can be extracted directly from the metric irrespective
of coordinates. The framework developed in this paper allows the
study of exact solutions in any frame without transformations.
Surprisingly little work has been done in non-comoving frames
\cite{com}. The invariant procedure developed is algorithmic and
suited to computer algebra projects such as GRDB \cite{computer}
where, given a metric, one might like to know if it is
necessarily, for example, a perfect fluid, or on the contrary,
incompatible with a perfect fluid. The paper is organized as
follows: In section II, we explore the zero flux condition and in
section III we derive explicit forms of the velocity field for
canonical coordinate types. In section IV we explore the
phenomenology of the non-conducting fluid source showing that
degenerate cases exist.  Examples that illustrate the power of the
results obtained are given in section V. Section VI is a summary.

\section{Zero Flux}
We consider warped product spacetimes of class $B_{1}$
\cite{santo,carot}. These can be written in the form
\begin{equation}
ds^2_{\mathcal{M}}=ds^2_{\Sigma_{1}}(x^1,x^2)+C(x^{\alpha})ds_{\Sigma_{2}}^{2}(x^3,x^4)\label{gmetric}
\end{equation}
where $C(x^{\alpha})=r(x^1,x^2)^2 w(x^3,x^4)^2$,
$sig(\Sigma_{1})=0$ and
 $sig(\Sigma_{2})=2 \epsilon$ $(\epsilon=\pm1)$.
Although very special, these spaces include many of interest, for
example, \emph{all} spherical, plane, and hyperbolic spacetimes.
We write
\begin{equation}
ds^{2}_{\Sigma_{1}}=a(dx^1)^{2}+2bdx^1dx^2+c(dx^2)^{2},
\label{metric}
\end{equation}
with $a,b$ and $c$ functions of $(x^1,x^2)$ only. A congruence of
unit timelike vectors (a ``flow" in what follows)
$u^{\alpha}=(u^1,u^2,0,0)$ have an associated unit normal field
$n^{\alpha}$ (in the tangent space of $\Sigma_{1}$) satisfying
$n_{\alpha}u^{\alpha}=0, n_{\alpha}n^{\alpha}=1$ \cite{notation}.
It follows that $n_{\alpha}=\psi(x^1,x^2)(u^2,-u^1,0,0)$ ($\psi$ a
normalization factor). The timelike condition on $u^{\alpha}$ is
\begin{equation}
-1=a(u^1)^{2}+2bu^1u^2+c(u^2)^{2} \label{timelike}
\end{equation}
and $(\mathcal{M},\bf{g})$ is time orientated by the restriction
\begin{equation}
u^1>0. \label{orientation}
\end{equation}

The condition
\begin{equation}
G_{\alpha}^{\beta}u^{\alpha}n_{\beta}=0, \label{flux}
\end{equation}
where $G_{\alpha}^{\beta}$ is the Einstein tensor, can be written
as
\begin{equation}
Au^1u^2-B(u^1)^2+C(u^2)^2=0, \label{flux1}
\end{equation}
where
\begin{equation}
A \equiv G_1^1-G_2^2, \;\;\; B \equiv G_1^2, \;\;\; C \equiv
G_2^1. \label{ABC}
\end{equation}
Condition (\ref{flux}) is the defining criterion for the flows
considered here. In general, $A$, $B$ and $C$ are functions of
$x^1$ through $x^4$, assumed not to vanish simultaneously.
Equations (\ref{timelike}), (\ref{orientation}) and (\ref{flux1})
determine the flow uniquely. The flow need not exist. For example,
if $A=C=0$ and $B\neq0$ then there exists no such flow. The assumption
of ``comoving coordinates" ($u^2=0$ in the present notation)
imposes restrictions on the coordinatization of $\Sigma_{1}$, but it
is already clear from (\ref{flux1}) that the existence of such a
flow requires $B=0$.

\section{$u^{\alpha}$ for Canonical Representations of $\Sigma_{1}$}
There are four distinct canonical types of coordinates (not
specific coordinates) that can be used to represent $\Sigma_{1}$. In
what follows we write out $u^{\alpha}$ in each of these cases. We
do not write out quantities that follow algorithmically from
$u^{\alpha}$.
\subsection{Kruskal-Szekeres ($a=c=0$)}
The distinguishing characteristic here is
\begin{equation}
A = 0. \label{null}
\end{equation}
We can always choose $b<0$ so that
\begin{equation}
u^2=-\frac{1}{2bu^1}>0, \label{nullu2}
\end{equation}
and so it follows from (\ref{flux1}) that
\begin{equation}
u^1=\sqrt[4]{\frac{C}{4Bb^2}}. \label{nullu1}
\end{equation}
\subsection{Diagonal ($b=0$)}
Here $a<0$, $c>0$ and
\begin{equation}
Bc=Ca. \label{diagonal}
\end{equation}
With
\begin{equation}
\mathcal{D} \equiv A^2a+4B^2c \label{D}
\end{equation}
it follows that $Da>0$,
\begin{equation}
u^1=\sqrt{\frac{1}{2}(\frac{-1}{a}+\sqrt{\frac{A^2}{\mathcal{D}a}})},
\label{diagonalu1}
\end{equation}
and
\begin{equation}
(u^2)^2=\frac{1}{2c}(-1+\sqrt{\frac{A^2a^2}{\mathcal{D}a}}).
\label{diagonalu2}
\end{equation}
The requirement of ``comoving" coordinates is $u^2=0
\Leftrightarrow B=C=0$.

\subsection{Bondi ($c=0$ or $a=0$)}
It is sufficient to consider the case $c=0$. Now $b \neq 0$ and
for $b>0$ $x^1$ is an ``advanced" time and for $b<0$ $x^1$ is a
``retarded" time. Now
\begin{equation}
Ab=Ca. \label{bondi}
\end{equation}
With
\begin{equation}
\mathcal{E} \equiv Aa^2+4Bab \label{E}
\end{equation}
it follows that $sign(\mathcal{E})=sign(A)$,
\begin{equation}
u^1=\sqrt[4]{\frac{A}{\mathcal{E}}}, \label{bondiu1}
\end{equation}
and
\begin{equation}
u^2=-\frac{1}{2b}(\sqrt[4]{\frac{\mathcal{E}}{A}})(1+a\sqrt{\frac{A}{\mathcal{E}}}).
\label{bondiu2}
\end{equation}
The condition of ``comoving" coordinates $u^2=0$ requires $a<0$
and $B=0$. Conversely, if $B=0$ then $u^2=0$ for
$a<0$ and $u^2=-\sqrt{a}/b$ for $a>0$.
\subsection{$a\not=0$, $b\not=0$ and $c\not=0$}
With
\begin{equation}
\mathcal{R} \equiv A^2c-2AbC+2CBc+2C^2a
\end{equation}
and
\begin{equation}
\mathcal{S} \equiv 2AbcB-2AbCa+B^2c^2-4BCb^2+2BcCa+C^2a^2+caA^2,
\end{equation}
it follows that
\begin{equation}
(u^1)^2=\frac{-\mathcal{R}+sign(b)\sqrt{\mathcal{R}^2-4\mathcal{S}C^2}}{2\mathcal{S}}
\label{generalu1}
\end{equation}
and
\begin{equation}
u^2=\frac{-bu^1+sign(b)\sqrt{(bu^1)^2-c(a(u^1)^2+1)}}{c}. \label{generalu2}
\end{equation}
The condition of ``comoving" coordinates $u^2=0$ requires the same
conditions as in the previous case.
\section{Phenomenology}
The discussion in this section requires no specification of
coordinates on $\Sigma_{1}$.
\subsection{Decomposition}
Whereas condition (\ref{flux}) determines the flow on $\Sigma_{1}$, it
is of interest to know when the flow reduces to that of a perfect
fluid (isotropic pressure (including bulk stress) and zero shear
stress), or conversely, to be able to state that a given spacetime
can not be compatible with a perfect fluid source. The familiar case
is entirely obvious: If the Einstein tensor is
diagonal with three equal components, then the metric is
consistent with a perfect fluid in comoving coordinates. Such a
circumstance is, however, a property of the spacetime and the
coordinates in which it is exhibited, more general cases are explored
in this section. With $u^{\alpha}$ known, it
is possible to reconstruct phenomenological parameters associated
with a decomposition of the energy-momentum tensor directly from
the metric. This decomposition is not always unique.

A phenomenological fluid interpretation of (\ref{flux}), by way of
Einstein's equations, follows from the well known Eckart relation
\cite{ecart} which gives
\begin{equation}
8 \pi \kappa n^\alpha(\nabla_\alpha\textit{T} + \textit{T} u^\beta
\nabla _\beta u_\alpha)=0
\end{equation}
where $\kappa$ is the thermal conductivity and $T$ the temperature
profile. The flows considered here can therefore be considered
non-conducting ($\kappa=0$) setting aside contrived functions $T$.
The energy-momentum tensor is decomposed here in the form
\cite{flow}
\begin{equation}
T^{\alpha}_{\beta}=\rho
u^{\alpha}u_{\beta}+p_{1}n^{\alpha}n_{\beta}
+p_{2}\delta^{\alpha}_{\beta} +
p_{2}(u^{\alpha}u_{\beta}-n^{\alpha}n_{\beta})-2\eta
\sigma^{\alpha}_{\beta}, \label{imperfect}
\end{equation}
where $\sigma^{\alpha}_{\beta}$ is the shear associated with
$u^{\alpha}$ and $\eta$ the phenomenological shear viscosity.
It follows from (\ref{gmetric}) that $G_{3}^{3}=G_{4}^{4}$ for
the spacetimes considered in this paper.
With $G_{3}^{3}=G_{4}^{4}$ it follows from (\ref{imperfect}) that
either $\sigma_{3}^{3}=\sigma_{4}^{4}$ or we must set $\eta \equiv
0$. Note that (\ref{imperfect}) distinguishes the shear stress
from an anisotropic pressure. These are sometimes combined. For
example, in the comoving frame in a spherically symmetric
spacetime, it follows that
$T^{\alpha}_{\beta}=diag(\rho,P1,P2,P2)$ where $P1=p_1-2 \eta
\sigma_2^2$ and $P2=p_2-2 \eta \sigma_3^3$. Such a combination is
not, in general, possible outside the comoving frame. Since
anisotropic pressures do not arise solely due to shear stresses
(\textit{e.g.} in a static spherically symmetric spacetime,
$\sigma_{\alpha}^{\beta}=0$ but $p_1 \neq p_2$ in general), the
full decomposition (\ref{imperfect}) is used here.
\subsection{Systems of Equations}
We set up systems of equations to be solved simultaneously for the
functions $(\rho,p_1,p_2,\eta)$ in terms of scalars that follow
algorithmically from the metric. It is not possible to build more
than three independent equations in an attempt to solve for
$(\rho,p_1,p_2,\eta)$ since, with (\ref{flux}), there are only
three independent scalars that can be constructed from the set
$(G_{\alpha}^{\beta}, u^{\alpha}, n_{\alpha})$ \cite{santo}.

\subsubsection{General Spacetimes}

 To proceed in a manifestly invariant way
we construct scalars from the set $(G_{\alpha}^{\beta},
u^{\alpha}, n_{\alpha})$ that are linear in $G_{\alpha}^{\beta}$.
These are:
$G_{\alpha}^{\beta}u^{\alpha}n_{\beta}$ (used in condition
(\ref{flux})), $G_{\alpha}^{\beta}u^{\alpha}u_{\beta}$,
$G_{\alpha}^{\beta}n^{\alpha}n_{\beta}$ and $
G_{\alpha}^{\alpha}\equiv G$. With (\ref{imperfect}) it follows
that
\begin{equation}
G=8 \pi ( -\rho+p_1+2p_2),\label{trace}
\end{equation}
and
\begin{equation}
G_{\alpha}^{\beta}u^{\alpha}u_{\beta} \equiv G1=8 \pi \rho.
\label{energy}
\end{equation}
In all cases we take (\ref{energy}) as the definition of $\rho$.
Further,
\begin{equation}
G_{\alpha}^{\beta}n^{\alpha}n_{\beta} \equiv G2=8 \pi(p_1-2 \eta
\Delta) \label{pressure}
\end{equation}
where
\begin{equation}
\Delta\equiv\sigma_{\alpha}^{\beta}n^{\alpha}n_{\beta}.\label{Delta}
\end{equation}
Rearrangement of (\ref{trace}), (\ref{energy}) and
(\ref{pressure}) gives
\begin{equation}
48 \pi \eta \Delta = G + G1-3G2+16 \pi P \label{eta}
\end{equation}
where $P \equiv p_{1}-p_{2}$.  If $\sigma_{\alpha}^{\beta}=0$ or
$\eta \equiv 0$ or $P \equiv 0$ then (\ref{trace}), (\ref{energy})
and (\ref{pressure}) form a complete set of equations. More
generally, however, if we attempt to solve for the complete set of
parameters $(\rho,p_1,p_2,\eta)$ four equations are needed
 \cite{components}.
For the class of spacetimes considered here,
it was shown in \cite{santo} that higher-order invariants are not
independent from the linear ones.
Therefore, use of  higher-order invariants will not break the
degeneracy but merely generate new syzygies (algebraic identities
amongst invariantly defined quantities).
\subsubsection{Restricted Spacetimes}
From (\ref{imperfect}) and Einstein's equations it follows that
for the spacetimes $(\mathcal{M},\bf{g})$ considered here
\begin{equation}
\widetilde{G}_{2}^{1}\widetilde{G}_{1}^{2}-(\widetilde{G}_{1}^{1}
-\widetilde{G}_{3}^{3})(\widetilde{G}_{2}^{2}-\widetilde{G}_{3}^{3})
\equiv \widetilde{G5}=(8 \pi)^2 P(\rho+p_{2}), \label{walker}
\end{equation}
where $ \widetilde{G}^{\alpha}_{\beta}= G^{\alpha}_{\beta}+16 \pi
\eta \sigma^{\alpha}_{\beta}$. Although (\ref{walker}) holds in
every $(\mathcal{M},\bf{g})$ without specific coordinates specified on
$\Sigma_{1}$, it is not manifestly invariant. Use of (\ref{walker})
merely generates  further (restricted) syzygies.

\subsection{Linear cases}
\subsubsection{$\sigma_{\alpha}^{\beta}=0$}
If $\sigma_{\alpha}^{\beta}=0$ and $\rho+p_2\neq0$ then equation
(\ref{walker}) with $P=0$ gives the Walker's
pressure isotropy condition \cite{walker}
\begin{equation}
G_{2}^{1}G_{1}^{2}=(G_{1}^{1}-G_{3}^{3})(G_{2}^{2}-G_{3}^{3})\label{walker1}
\end{equation}
which is here a necessary and sufficient condition for a perfect fluid. A
manifestly invariant condition follows from (\ref{trace}),
(\ref{energy}) and (\ref{pressure}) which give

\begin{equation}
p_1=\frac{G2}{8 \pi}, \label{p10}
\end{equation}
and
\begin{equation}
p_2=\frac{G+G1-G2}{16 \pi}. \label{p20}
\end{equation}
Clearly
\begin{equation}
G+G1=3G2\label{giso}
\end{equation}
is a necessary and sufficient condition for a perfect fluid
including the exceptional case $G+3G1=G2$ not covered by
(\ref{walker1}). Equation (\ref{giso}) is not sensitive to the
presence of a cosmological constant term in the Einstein field
equations and plays a central role in what follows \cite{alternate}

\bigskip

In all of what now follows up to section V we assume
$\sigma_{\alpha}^{\beta} \neq 0$.
\subsubsection{$\eta \equiv 0$}
The decomposition (\ref{imperfect}) is consistent with some
spacetimes if and only if $\eta \equiv 0$. Some examples of this
are shown in section \ref{examples}. If $\eta \equiv 0$ then
equations (\ref{p10}), (\ref{p20}) and (\ref{giso}) hold as in the
previous case.
\subsubsection{$p_1=p_2 \equiv p, \; \Delta \neq 0$}

Equations (\ref{trace}), (\ref{energy}) and (\ref{pressure}) now
give

\begin{equation}
p=\frac{G+G1}{24 \pi} \label{pc2}
\end{equation}
and
\begin{equation}
\eta=\frac{G+G1-3G2}{48 \pi \Delta} \label{etac2}
\end{equation}
so that (\ref{giso}) is once again a necessary and sufficient
condition for a perfect fluid. The case $\Delta=0$ is equivalent
to the case $\sigma_{\alpha}^{\beta}=0$.
\subsubsection{$p_1 \neq p_2 , \; \Delta \neq 0$}
Equations (\ref{trace}), (\ref{energy}) and (\ref{pressure}) now
give
\begin{equation}
p1=\frac{G2}{8 \pi}+2 \eta \Delta, \label{p1s}
\end{equation}
and
\begin{equation}
p2=\frac{G+G1-G2}{16 \pi}-\eta \Delta, \label{p2s}
\end{equation}
where $\eta$ is arbitrary. If we set $\eta\equiv 0$ then the
condition (\ref{giso}) is a necessary and sufficient condition for
a perfect fluid. For other choices of $\eta$ the fluid is
imperfect.

\bigskip
\subsection{Non-uniqueness of the source}
In the last case above, it is not
possible to solve for a unique set $(\rho,p_1,p_2,\eta)$ as
only three invariants are independent for the type of spacetimes
considered in this paper.
 Another way to see this
directly is to observe that substitution of the expressions for
$\rho$, $p_1$ and $p_2$ as given by (\ref{energy}), (\ref{p1s})
and (\ref{p2s}) into the energy-momentum tensor (\ref{imperfect})
and multiplication by $8\pi$ reproduces the Einstein tensor
\cite{lake2002a}. This of course all derives from the fact that in
a canonical frame there are at most three independent components
of the Einstein tensor for the spacetimes considered.
As we show in section \ref{examples}, the
application of the framework to some given spacetimes known to
represent perfect fluid solutions but where $\Delta \neq 0$ shows
that there are other imperfect fluid sources possible. This is
always the case when the perfect fluid condition (\ref{giso})
holds and $\Delta \neq 0$. For $\eta \equiv 0$ the fluid is
perfect, for other choices of $\eta$ the fluid is imperfect. For
example, the Lema\^{\i}tre-Tolman-Bondi metric (``dust'') is given
by \cite{lemaitre}\cite{krasinski}
\begin{equation}
ds^2_{\mathcal{M}}=-(dt)^2+\frac{(R^{'}(t,r))^2 (dr)^2}{1+f(r)}
+R(t,r)^2 d\Omega^2, \label{tolman}
\end{equation}
along with the constraints
\begin{equation}
\dot{R}(t,r)=\sqrt {2\,{\frac {m( r) }{R( t,r ) }}+f ( r) },
\end{equation}
\begin{equation}
\ddot{R}(t,r)=-{\frac {m ( r) }{ R ^{ 2}( t,r)}},
\end{equation}
\begin{equation}
 \ddot{R}^{'}(t,r)=-{\frac {m ^{'}( r ) }{  R ^{2}( t,r )
 }}+2\,{\frac {m ( r ) R^{'}( t,r) }{  R ^{3}( t,r
 )   }},
 \end{equation}
 and
% \begin{widetext}
 \begin{equation}
 \dot{R}^{'}(t,r)=
 {\frac {2\, m ^{'}( r) R
( r,t ) -2\,m ( r )R^{'} ( r,t ) +  f^{'} ( r
 )    R ^{2}( r,t )   }{2R
 ( r,t ) \sqrt { ( 2\,m ( r ) +f ( r
 ) R ( r,t )) R ( r,t) }}}
, \end{equation}
%\end{widetext}
where $^{'} \equiv \frac{\partial}{\partial r}$ and $^{.} \equiv
\frac{\partial}{\partial t}$. Condition (\ref{giso}) holds
so the source is consistent with (but not
necessarily) a perfect fluid, in fact simply dust (since
$G2=0=G+G1$) with
\begin{equation} \rho={\frac {{ }m^{'} ( r ) }{ 4 \pi R^2 (
r,t ) \,{ }R^{'} ( r, t ) }}.
\end{equation}
The metric (\ref{tolman}) (with the given constraints) is also
consistent with an imperfect fluid with $p_1=2 \eta \Delta$,
$p_2=- \eta \Delta$, $\eta$ arbitrary and
\begin{widetext}
\begin{eqnarray}
\Delta=-{\frac {-2\,  m^{'}(r,t)  R
 ( r,t ) +6\,m ( r )
R^{'}(r,t) - f ^{'}( r
 )   R^{2} ( r,t )   +2\,
 R ^{'}( r,t )  f
( r ) R( r,t )}{ 3R^{'} ( r,t )  ) R ^{3/2}( r,t) \sqrt {
  2\,m ( r ) +f ( r ) R ( r,t
 ) }}}
\end{eqnarray}
\end{widetext}
as is easily verified. The fact that the Lema\^{\i}tre metric need
not be considered as dust is not a new result \cite{tolmans}. Here, the
degenerate case ($\Delta \ne 0$) is shown to be generic.

\section{Examples} \label{examples}
We provide here examples in various coordinate types in order to illustrate
the results obtained.
\subsection{Kruskal-Szekeres coordinates}
Aside from the Kruskal-Szekeres metric, little use has been made
of double null coordinates. The example used here is the
Einstein-de Sitter universe \cite{laapois}
\begin{equation}
ds^2_{\mathcal{M}}=\mathcal{C}^2(u+v)^4(-dudv+\frac{(u-v)^2}{4}d\Omega^2),
\end{equation}
where $\mathcal{C}$ is a constant and $d\Omega^2$ is the metric of
a unit sphere. It follows that $u^{\alpha}=
\frac{1}{\mathcal{C}(u+v)^2}(1,1,0,0)$,
$\sigma_{\alpha}^{\beta}=0$ and (\ref{giso}) holds so that the
fluid is necessarily perfect (in fact simply dust since $G2=0=G+G1$).
\subsection{Bondi coordinates}
The Bondi metric \cite{bondip}
\begin{equation}
ds^2_{\mathcal{M}}=c^2(w,r)f(w,r)(dw)^2 \pm 2 c(w,r) dw dr +r^2
d\Omega^2 \label{bondim}
\end{equation}
in advanced ($+$) or retarded ($-$) $w$ has $A=C=0, B\neq0$ for
$\frac{\partial c}{\partial r}=0$ and $\frac{\partial f}{\partial
w}\neq0$ and so there is no non-conducting fluid source of
(\ref{bondim}) under these conditions. The Vaidya metric
\cite{vaidya} (corresponding to a null flux) provides a familiar
example. The metric \cite{davidson}
\begin{equation}
ds^2_{\mathcal{M}}=-2H(u,r)(du)^2-2dudr+ur^{2n}((dx)^2+(dy)^2),\label{davidson}
\end{equation}
where $H(u,r)=(r/u+kr^{m}u^{(2-m)/(m-1)})/2>0$ and $n=m(m-1)/2$ is
necessarily comoving. Moreover, $\sigma_{\alpha}^{\beta} \neq 0$
and (\ref{giso}) holds so the source
is consistent with (but not necessarily) a perfect fluid. If
we set $\eta\equiv 0$ then the fluid is perfect but for other choices
of $\eta$ the fluid is imperfect.

\subsection{Comoving diagonal coordinates}
The next examples are in diagonal coordinates and have $B=C=0$ so
that $u^{\alpha}$ is necessarily comoving.
\subsubsection{$\sigma_{\alpha}^{\beta}=0$}

The Robertson-Walker metric
\begin{equation}
ds^2_{\mathcal{M}}=-(dt)^2+\frac{a(t)^2 (dr)^2}{1-kr^2} +r^2
d\Omega^2 \label{rw}
\end{equation}
gives $\sigma_{\alpha}^{\beta}=0$ and (\ref{giso}) holds. It
follows that the fluid is necessarily perfect \cite{rwalt}.
Equations (\ref{energy}) and (\ref{pressure}) reproduce
Friedmann's equations. In contrast, the Kantowski-Sachs metric \cite{ksm}
\begin{equation}
ds^2_{\mathcal{M}}=-(dt)^2+\frac{a(t)^2 (dr)^2}{1-kr^2} +
d\Omega^2 \label{ksachs}
\end{equation}
gives $\sigma_{\alpha}^{\beta}=0$ but (\ref{giso}) never holds. It
follows that the fluid can never be perfect.
For the general spherical static metric
\begin{equation}
ds^2_{\mathcal{M}}=-e^{2 \Phi(r)}(dt)^2+\frac{(dr)^2}{1-2m(r)/r}
+r^2 d\Omega^2, \label{static}
\end{equation}
whereas $\sigma_{\alpha}^{\beta}=0$, (\ref{giso}) does not in
general hold. The fluid has anisotropic pressure, the perfect
fluid being a special case \cite{delake}. The metric \cite{mitsen}
\begin{equation}
ds^2_{\mathcal{M}}=-(dt)^2+R(t)^2((dr)^2+sin(r)^2(dz)^2+f(r)^2(d\phi)^2)
\label{mitsen}
\end{equation}
with the constraint $2R\ddot{R}+(\dot{R})^2+1=0$ where $^{.}\equiv
\frac{\partial}{\partial t}$ has
$\sigma_{\alpha}^{\beta}=G_{\phi}^{\phi}=0$. The condition
$G_{z}^{z}=0$ gives $f(r)=cos(r+A)$ where $A$ is constant.
Condition (\ref{giso}) does not in general hold. Rather, $p_2=0$
but $p_1=0$ only for $A=0$.

\subsubsection{$\eta \equiv
0$} The metric \cite{ds}
\begin{eqnarray}
ds^2_{\mathcal{M}}&=&S(t)^{-2m}C(x)^{-2m-2}(-(dt)^2+(dx)^2)
\nonumber\\
&&+S(t)C(x)^{\alpha}(T(t)^n(dy)^2+T(t)^{-n}(dz)^2)
\end{eqnarray}
has $G_{x}^{x}=G_{y}^{y}=G_{z}^{z}$ and so is obviously a perfect
fluid in comoving coordinates. However, $\sigma_{y}^{y} \neq
\sigma_{z}^{z}$ and so the decomposition (\ref{imperfect}) holds
only for $\eta=0$. Condition (\ref{giso}) holds in agreement with
the obvious.
\subsection{Non-comoving diagonal coordinates}
Few examples are available in non-comoving coordinates. From the
pioneering work of McVittie and Wiltshire \cite{com} we note for
example, that their solution (6.12)
\begin{equation}
ds^2_{\mathcal{M}}=e^{2\beta(z)}(-(dt)^2+(d
\xi)^2+\xi^2 d\Omega^2) \label{mcv612}
\end{equation}
where $z=\epsilon (\xi^2-t^2)/\xi_0^2$, and $\beta(z)$ is an
undetermined function of z with $\beta_{zz}-\beta^{2}_z \ne 0$, has
$\sigma_{\alpha}^{\beta} = 0$ and the  condition (\ref{giso})
holds so the source is necessarily a perfect fluid. Their solution
(7.20)
\begin{equation}
ds^2_{\mathcal{M}}=exp(Ae^{z/L}+Bz/l-2 \epsilon L t)(-(dt)^2+(d
\omega)^2+d\Omega^2) \label{mcv}
\end{equation}
where $z=\omega+\epsilon t$, $\epsilon = \pm 1$ and $A, B$ and $L$
are constants, has $\sigma_{\alpha}^{\beta} \neq 0$. Condition
(\ref{giso}) holds so the source is consistent with a perfect
fluid (if $\eta \equiv 0$). Their solutions (6.21) and (8.11) also
have shear and (\ref{giso}) holds so the solutions are compatible
with a perfect fluid source. As with the Davidson metric
(\ref{davidson}), McVittie and Wiltshire solutions (6.21), (7.20)
and (8.11) are also consistent with an imperfect fluid with $\eta
\neq 0$.

From the more recent work of Senovilla and Vera \cite{com}, for example, their
solution (40)
\begin{eqnarray}
ds^2_{\mathcal{M}}&=&-(dt)^2+(dx)^2+\frac{cos^{1+2 \nu}(\mu
x)}{cosh^{2 \nu -1}(\mu t)}(dy)^2\nonumber\\
&&+\frac{cos^{1-2 \nu}(\mu
x)}{cosh^{-2 \nu -1}(\mu t)}(dz)^2\label{seno}
\end{eqnarray}
has $G_y^y=G_z^z$ but $\sigma_y^y \neq \sigma_z^z$ and so the
metric (\ref{seno}) is consistent with the decomposition
(\ref{imperfect}) only for $\eta \equiv 0$ (the same holds for their
solutions (38) and (41)). Equation (\ref{giso})
holds and so the metric (\ref{seno}) necessarily represents a
perfect fluid.

\section{Summary}

It has been shown that a flow (timelike congruence $u^{\alpha}$)
in any type $B_{1}$ warped product spacetime is uniquely and
algorithmically determined by the condition of zero flux
($G_{\alpha}^{\beta}u^{\alpha}n_{\beta}=0$). Explicit forms of
$u^{\alpha}$ have been written out for canonical representations.
With $u^{\alpha}$ known, a phenomenological interpretation of the
spacetime $(\mathcal{M},\bf{g})$ in terms of a non-conducting
fluid follows. The following cases are delineated
\begin{itemize}
\item{i. If $\eta \Delta = 0$
then the condition (\ref{giso})
is a necessary and
sufficient condition for a perfect fluid.}
\item{ii. If $p_1 \equiv p_2 \equiv p$ and $\Delta \not= 0$
then $p$ is given by (\ref{pc2}) and $\eta$ is given by (\ref{etac2}).
The condition (\ref{giso}) is a
necessary and sufficient condition for a perfect fluid.}
\item{iii. If $\Delta \not= 0$ then $p_1$ is given by
  (\ref{p1s}) and $p_2$ is given by (\ref{p2s}) and $\eta$ is a freely
  specified function. This is a generic degenerate case. If condition
  (\ref{giso}) holds then the
  fluid is compatible with a perfect fluid source. If $\eta \equiv 0$
  the fluid is perfect. For other choices of $\eta$ the fluid is imperfect.}
\end{itemize}
Furthermore, the derived covariant perfect fluid condition can be
used to study exact solutions of Einstein's equations as well as
deriving new families of solutions \cite{lake2002b}. Examples, in
various coordinates, including spacetimes with and without unique
interpretations have been provided. The
procedure developed has been implemented in a computer algebra
program described in the appendix.

\begin{acknowledgments}
We would like to thank the referee for useful comments.
This work was supported by a grant (to KL) from the Natural
Sciences and Engineering Research Council of Canada (NSERC) and by an
Ontario Graduate Scholarship and an NSERC PDF Fellowship (to
MI). Portions of this work were made possible by use of
\textit{GRTensorII} \cite{grt}.
\end{acknowledgments}

\begin{appendix}

\section{GRSource, a computer algebra program}\label{grsource}
\textit{GRSource} \cite{grs} runs under the system
\textit{GRTensorII} which in turn runs under the system Maple. The
program is called with a spacetime metric name as an input.
\textit{GRSource} calculates the velocity field and related
physical quantities. Moreover, it performs an algorithmic analysis
on the nature of the fluid. The results of the evaluated
quantities are displayed, followed by a summary report on the
analysis of the possible fluid sources for the input spacetime. An
example follows:

\begin{widetext}
\emptyline
{$>$ restart;}
\emptyline
We start a GRTensorII and GRSource session

\emptyline
{$>$ grtw();}
\begin{center}
\it{GRTensorII\ Version\ 1.80-pre2\ (R6)}

\it{Developed\ by\ Peter\ Musgrave,\ Denis\ Pollney\ and\
Kayll\ Lake}

\it{Copyright\ 1994-2003\ by\ the\ authors.}

\it{Latest\ version\ available\ from:\ http://grtensor.org}

\emptyline
\it{GRSource\ Package\ Version\ 1.00}

\it{Developed\ by\ Mustapha\ Ishak\ and\ Kayll\ Lake,\ (c)\
2002-2003}

\it{Help\ available\ via\ ?grsource}

\it{Usage:\ Load\ a\ metric\ and\ enter\ the\ command\
source(metricname);\ }
\end{center}

\emptyline \noindent $>$ qload(kantosachs);
\[
\mathit{Default\ spacetime}=\mathit{kantosachs}
\]
\[
\mathit{For\ the\ kantosachs\ spacetime:}
\]
\[
\mathit{Coordinates}\,:\,\mathrm{x}(\mathit{up})
\]
\[
\mathit{x\ }^{a}=[r, \,\theta , \,\phi , \,t]
\]
\[
\mathit{Signature\ }=2
\]
\[
\mathit{Line\ element}\,:\,\mathit{ds^2}
\]
\[
\mathit{\ ds}^{2}={\displaystyle \frac {\mathrm{a}(t)^{2}\,
\mathit{\ d}\,r^{\mathit{2\ }}}{1 - K\,r^{2}}}  + \mathrm{a}(t)^{
2}\,\mathit{\ d}\,\theta ^{\mathit{2\ }} + \mathrm{a}(t)^{2}\,
\mathrm{sin}(\theta )^{2}\,\mathit{\ d}\,\phi ^{\mathit{2\ }} -
\mathit{\ d}\,t^{\mathit{2\ }}
\]
\[
\mathit{Kantowski-Sachs\ metric,\ J.\ Math.\ Phys.\ 7,\ 443\
(1966)}
\]
$>$ source(kantosachs);

\emptyline
A general velocity field will be generated automatically from the
metric.

Please answer the following or enter exit anytime to stop the session.

\emptyline
{Enter the timelike  or null coordinate, for example: t; }{$>$ t;}

{Enter the spacelike or null coordinate, for example: r; }{$>$ r;}

{Enter the spacelike coordinate, for example: theta; }{$>$ theta;}

{Enter the spacelike coordinate, for example: phi; }{$>$ phi;}
\[
\mathit{For\ the\ kantosachs\ spacetime:}
\]
\[
\mathit{u}^{a}=[0, \,0, \,0, \,1]
\]
\[
\mathit{uSq}=-1
\]
\[
\mathit{flux}=\mathit{All\ components\ are\ zero}
\]

\[
\mathit{ShearTensor}^{a}\,{\ _{b}}=\mathit{All\ components\ are\
zero}
\]
\[
\mathit{Expansion\ scalar}\,:\,\mathit{expsc}
\]
\[
\mathit{Theta[u]}=3\,{\displaystyle \frac {{\frac {d}{dt}}\,
\mathrm{a}(t)}{\mathrm{a}(t)}}
\]
\[
\mathit{Delta}=\mathit{All\ components\ are\ zero}
\]
\[
\mathit{rho}={\displaystyle \frac {1}{8}} \,{\displaystyle \frac
{3\,({\frac {d}{dt}}\,\mathrm{a}(t))^{2} + 1}{\mathrm{a}(t)
^{2}\,\pi }}
\]
\[
\mathit{p1}= - {\displaystyle \frac {1}{8}} \,{\displaystyle
\frac {({\frac {d}{dt}}\,\mathrm{a}(t))^{2} + 2\,\mathrm{a}(t)\,(
{\frac {d^{2}}{dt^{2}}}\,\mathrm{a}(t)) + 1}{\mathrm{a}(t)^{2}\,
\pi }}
\]
\[
\mathit{p2}= - {\displaystyle \frac {1}{8}} \,{\displaystyle
\frac {({\frac {d}{dt}}\,\mathrm{a}(t))^{2} + 2\,\mathrm{a}(t)\,(
{\frac {d^{2}}{dt^{2}}}\,\mathrm{a}(t))}{\mathrm{a}(t)^{2}\,\pi }
}
\]
\[
\mathit{PFCondition}=2\,{\displaystyle \frac {1}{\mathrm{a}(t)^{2
}}}
\]
\[
\mathit{Report:}
\]
\[
\mathit{eta*Delta\ =\ 0\ but\ PFCondition\ was\ not\ simplified\
to\ zero,}
\]
\[
\mathit{the\ fluid\ has\ anisotropic\ pressure.}
\]
\[
\mathit{Further\ simplifications\ can\ be\ applied\ to\ the\
objects\ calculated}
\]
\[
\mathit{using\ the\ commands\ gralter()\ and\ grdisplay().}
\]
\end{widetext}
\end{appendix}


\begin{thebibliography}{}
\bibitem{solutions}Any spacetime can be associated with an ``exact
  solution" to Einstein's  equations, but
only those which are ``physically reasonable" are worthy of
consideration. Whereas the term ``exact solution" has no precise
definition, the concept is well understood \cite{kshm}
\cite{krasinski}. The concept of ``physically reasonable" is
somewhat more slippery and it is the responsibility of the author
of any new exact solution to justify its significance.
\bibitem{flow} We are interested in fluid flows alone here and ignore
contributions from a null flux, and from scalar and
electromagnetic fields.
\bibitem{components} Component expressions follow
immediately from (\ref{imperfect}). For example, $G_{3}^{3}=8
\pi(p_2-2\eta \sigma_{3}^{3})$ could be used.
 These component expressions fail to solve for
a unique set $(\rho,p_1,p_2,\eta)$. Furtheremore, the approach used in
this section is manifestly invariant.
\bibitem{alternate} Equation (\ref{giso}) is linear in $G_{\alpha}^{\beta}$.
Alternate forms can be derived if one considers higher-order invariants
built only from the Einstein tensor or the trace-free Ricci tensor
($S^{\alpha}_{\beta}\equiv R^{\alpha}_{\beta} - \frac{R}{4}
\delta^{\alpha}_{\beta} $).
An example of such an equation is $3 R2^2-R1^3=0$ where
$R1\equiv \frac{1}{4}S^{\alpha}_{\beta}S^{\beta}_{\alpha}$ and
$R2\equiv -\frac{1}{8}S^{\alpha}_{\beta}S^{\beta}_{\gamma}S^{\gamma}_{\alpha}$.
The velocity and the normal vector fields do not enter in
such a relation,
however one needs to verify
the existence of the fluid flow (as explained earlier, the flow need
not exist).
\bibitem{grs}GRSource program is freely distributed from the authors.
\end{thebibliography}
\end{document}